\documentclass[10pt]{revtex4}
\usepackage{amssymb}                                       
\usepackage{latexsym}
\usepackage{epsfig}
\begin{document}

\title{Generalized ghost dark energy in Brans-Dicke theory}

\author{A. Sheykhi $^{1,2}$ \footnote{asheykhi@shirazu.ac.ir}, E. Ebrahimi $^{2,3}$ \footnote{eebrahimi@uk.ac.ir} and Y. Yousefi $^{3}$ }
\address{$^1$ Physics Department and Biruni Observatory, College of
Sciences, Shiraz University, Shiraz 71454, Iran\\
$^2$  Research Institute for Astronomy and Astrophysics of Maragha
         (RIAAM), P.O. Box 55134-441, Maragha, Iran\\
$^3$ Department of Physics, Shahid Bahonar University, PO Box
76175, Kerman, Iran }

\begin{abstract}
It was argued that the vacuum energy of the Veneziano ghost field
of QCD, in a time-dependent background, can be written in the
general form, $H + O(H^2)$, where $H$ is the Hubble parameter.
Based on this, a phenomenological dark energy model whose energy
density is of the form $\rho=\alpha H+\beta H^{2}$ was recently
proposed to explain the dark energy dominated universe. In this
paper, we investigate this generalized ghost dark energy model in
the setup of Brans-Dicke cosmology. We study the cosmological
implications of this model. In particular, we obtain the equation
of state and the deceleration parameters and a differential
equation governing the evolution of this dark energy model. It is
shown that  the equation of state parameter of the generalized
ghost dark energy can cross the phantom line ($w_D=-1$) in some
range of the parameters spaces.

\end{abstract}

 \maketitle

\section{Introduction}
Nowadays, it is a general belief that our  Universe is currently
experiencing a phase of accelerated expansion. Various
cosmological observations confirm this acceleration. The first
significant evidence was given from  measurements of type Ia
supernovaes [SNeIa] \cite{riess,snia2}. These results have been
confirmed repeatedly by several other observations such as
measurement of the anisotropies of the cosmic microwave background
(CMB), spectrum by the Wilkinson Microwave Anisotropy Probe (WMAP)
\cite{wmap}, and by the measurement of the baryon acoustic
oscillations (BAO) in the Sloan Sky Digital Survey (SDSS) luminous
galaxy sample \cite{tegmark}.

To explain such a phase of acceleration in the framework of
Einstein gravity we need to assume that the universe is filled
with an unknown type of energy component which is called dark
energy (DE). This component of energy has a negative equation of
state parameter (EoS) $w={p}/{\rho}<-\frac{1}{3}$ and is
responsible for such an acceleration. One main task for the
theoretical physicists is to identify the nature of such DE. The
simplest candidate is the famous cosmological constant with
$w=-1$, which is still one of the best, among various models, in
agreement with observations. However, this candidate suffers the
so called fine-tuning and the coincidence problems \cite{copsami}.
Further observations favor alternatives whose EoS parameter change
with time. Simplest example of this class is scalar fields which
have time varying EoS parameters. An incomplete list of the scalar
filed or time varying EoS parameter models can be found in
\cite{wetter,ratra,chiba,armend1,armend2,cald,feng, tachyon,
hol,age}. In addition to the DE approach, one can also explain the
late time acceleration of the universe with modified gravity
\cite{arkani} or inhomogeneous cosmology \cite{inhomo}. In these
approaches one assumes that the underlying theory of gravity
should be modified in such a way that the acceleration of the
universe expansion can be derived naturally from the theory.

Since any new model of DE has many unknown features and can lead
new problems in the literature, our prior choice is to handle the
DE problem without introducing new degrees of freedom beyond what
are already known. Recently one class of such models has been
attracted a lot of attentions, the so called "ghost dark energy"
(GDE). In this approach the Veneziano ghost field is responsible
for the recent cosmic acceleration. The Veneziano ghost field was
proposed to resolve the U(1) problem in QCD. The U(1) problem is
that the Lagrangian of QCD has, in the massless limit, a global
chiral U(1) symmetry, which does not seem to be reflected in the
spectrum of light pseudoscalar mesons. The ghost field has no
contribution in the vacuum energy in Minkowski spacetime but in
the time dependent or non-flat background the ghost filed has a
non vanishing contribution to the vacuum energy proportional to
$\Lambda_{\rm QCD}^3 H$, where $H$ is the Hubble parameter and
$\Lambda_{\rm QCD}^3$ is QCD mass scale \cite{Ohta}. In GDE model
the vacuum energy of the ghost field can be taken as a dynamical
cosmological constant \cite{Urban,Ohta}. Different features of GDE
have been explored in ample details
\cite{CaiGhost,shemov,Ebr,ebrins}. In \cite{zhi}, the author
discussed that the contribution of the Veneziano QCD ghost field
to the vacuum energy is not exactly of order $H$ and a subleading
term $H^2$ appears due to the fact that the vacuum expectation
value of the energy-momentum tensor is conserved in isolation
\cite{mig}. It was argued that the vacuum energy of the ghost
field can be written as $H+O(H^2)$, where the subleading term
$H^2$ in the GDE model might play a crucial role in the early
stage of  the universe evolution, acting as the early DE
\cite{CaiGhost2}. This term can also lead to a better agreement
with observations \cite{CaiGhost2,ebrsheyggde}.

On the other hand, in recent years, scalar tensor theories have
been reconsidered extensively. The reason comes from the fact that
scalar fields appear in different branches of theoretical physics
as a consistency condition. For example, the low energy limit of
the string theory leads to introducing a scalar degree of freedom.
One important example of the scalar tensor theories is the
Brans-Dicke (BD) theory of gravity which was presented by Brans
and Dicke in 1961 to incorporate the Mach's principle in the
Einstein's theory of gravity \cite{BD}. This theory also passed
the observational tests in the solar system domain
\cite{bertotti}. Since the generalized GDE (GGDE) model have a
dynamic behavior it is more reasonable to consider this model in a
dynamical framework such as BD theory. It was shown that some
features of original GDE in BD cosmology differ from Einstein's
gravity. For example while the original GDE is instable in all
range of the parameter spaces in standard cosmology \cite{ebrins},
it leads to a stable phase in BD theory \cite{saaidi}. All the
above reasons motivate us to investigate the GGDE model with
subleading term $H^2$ in the framework of BD theory.

This paper is organized as follows. In the next section, we review
the GGDE model in standard cosmology. In section \ref{EBDE in BD},
we extend the study to the framework of BD cosmology. We finish
with closing remarks in section \ref{sum}.
\section{Generalized Ghost dark energy Model}\label{EGDE}
For flat FRW universe filled with GDE and pressureless dark
matter, the first Friedmann equation may be written as
\begin{equation}\label{Fr1}
H^{2}=\frac{1}{3M_{p}^{2}}(\rho_{M}+\rho_{D}),
\end{equation}
where $\rho_m$ and $\rho_D$ are, respectively, the energy
densities of pressureless matter and ghost dark energy. The
generalized ghost energy density is \cite{CaiGhost2}
\begin{equation}\label{rhoD}
\rho_{D}=\alpha H+\beta H^{2}\label{d},
\end{equation}
where $\alpha$ is a constant with dimension $[energy]^3$, roughly
of order $\Lambda_{\rm QCD}^3$ and $\Lambda_{\rm QCD}\approx
100MeV$ is QCD mass scale, and $\beta$ is another constant with
dimension $[energy]^2$. We define the dimensionless density
parameters as usual,
\begin{equation}\label{Omega}
\Omega_m=\frac{\rho_m}{\rho_{cr}},\ \ \
\Omega_D=\frac{\rho_D}{\rho_{cr}}=\frac{\alpha+\beta H}{3M_p^2 H},
\end{equation}
where the critical energy density is $\rho_{cr}={3H^2 M_p^2}$.
Thus, the Friedmann equation can be rewritten as
$\Omega_m+\Omega_D=1$. The conservation equations read
\begin{eqnarray}
\dot\rho_m+3H\rho_m&=&0,\label{consm}\\
\dot\rho_D+3H\rho_D(1+w_D)&=&0\label{consd}.
\end{eqnarray}
Integrating Eq. ($\ref{consm}$) , we find
\begin{eqnarray}
\rho_m=\rho_{m0}(1+z)^3,
\end{eqnarray}
where $z=1/a-1$ is the redshift function. Thus, Friedmann equation
(\ref{Fr1}) can be written
\begin{equation}\label{Fr2}
H^{2} \Gamma-\alpha H = \rho_{m0}(1+z)^3,
\end{equation}
where $\Gamma=3M_p^2-\beta$. Solving the above equation we find
\begin{equation}\label{Fr3}
H(z)=\frac{1}{2\Gamma}\left(\alpha\pm \sqrt{\alpha^2+4\Gamma
\rho_{m0}(1+z)^3} \right).
\end{equation}
Using the fact that $\rho_{m0}=3M_p^2 H^2_{0}
\Omega_{m0}=(\Gamma+\beta)H^2_{0} \Omega_{m0}$, the above equation
can be further rewritten as
\begin{equation}\label{Fr4}
H(z)=\frac{1}{2\Gamma}\left(\alpha\pm \sqrt{\alpha^2+4\Gamma
(\Gamma+\beta) H^2_{0} \Omega_{m0}(1+z)^3} \right).
\end{equation}
Using the energy density ratio relation $u=\rho_{m}/\rho_{D}$, the
Friedmann equation can be also written as
\begin{eqnarray}
3M_{p}^{2}H^{2}=(1+u)\rho_{D}.\label{I}
\end{eqnarray}
Taking the time derivative of relation (\ref{d}) and using the
Friedmann equation we find $ \dot{\rho}_{D}=(\alpha+2\beta H)
\dot{H}$. Also, from the Friedmann equation as well as continuity
equations we have
\begin{eqnarray} \label{dotH}
\dot{H}=-\frac{1}{2M_{P}^{2}}(1+u+w_{D})\rho_{D}.
\end{eqnarray}
Substituting $ \dot{\rho}_{D}$ into Eq. (\ref{consd}), after some
simplifications, we find the EoS parameter of the GGDE as
\begin{eqnarray}
w_{D}=-\frac{1}{(2-\Omega_{D})+2\xi H(1-\Omega_{D})}, \label{wD1}
\end{eqnarray}
where $\xi={\beta}/{\alpha}$. It is easy to see that at the early
time where $\Omega_D\ll 1$ we have
\begin{eqnarray}
w_{D}=-\frac{1}{2}\left(\frac{1}{1+\xi H}\right), \label{wDe}
\end{eqnarray}
while at the late time where $\Omega_D\rightarrow 1$ the GGDE
mimics a cosmological constant, namely $w_D= -1$. This behavior is
similar to the original GDE \cite{shemov}. When $\xi=\beta=0$, one
recovers the EoS parameter of the original GDE \cite{shemov}
\begin{eqnarray}
w_{D}=-\frac{1}{2-\Omega_{D}}.
\end{eqnarray}
Also, the deceleration parameter is obtained as
\begin{eqnarray}
q=-1-\frac{\dot{H}}{H^{2}}=-1+\frac{3}{2}\Omega_{D}(1+u+\omega_{D}),\label{qq1}
\end{eqnarray}
where we have used Eq. (\ref{dotH}) and relation $\rho_D=3M_p^2
H^2$.  Substituting $w_D$ from (\ref{wD1}), we can further
simplify $q$ as
\begin{eqnarray}
q=\frac{1}{2}-\frac{3}{2}\Omega_{D}\left[2-\Omega_{D}+2\xi
H\left(1-\Omega_{D}\right)\right]^{-1}.\label{qq2}
\end{eqnarray}
At the late time where the dark energy dominates
($\Omega_D\rightarrow 1$) we have $q=-1$.  Taking the time
derivative of $\Omega_D$ in Eq. (\ref{Omega}), we find
\begin{eqnarray}
\dot{\Omega}_{D}=-\left(\frac{\alpha}{3M_{P}^{2}}\right)\frac{\dot{H}}{H^{2}}=\frac{\Omega_{D}H}{1+\xi
H}\left(1+q\right)
\end{eqnarray}
Using relation ${\dot{\Omega}_D}= H \frac{d\Omega_D}{d\ln a}$, we
obtain
\begin{eqnarray}
\frac{d\Omega_{D}}{d\ln a}&=&\frac{\Omega_{D}}{1+\xi
H}(1+q) \\
&&=\frac{3}{2}\left(\frac{\Omega_{D}}{1+\xi H}\right)\left[
1-\frac{\Omega_{D}}{(2-\Omega_{D})+2\xi
H(1-\Omega_{D})}\right].\label{dlna}
\end{eqnarray}
This is the equation of motion governing the evolution of GGDE.
The evolution of $H$ in Eqs. (\ref{qq2}) and (\ref{dlna}) is given
by Eq. (\ref{Fr4}).
\section{Generalized ghost dark energy in BD cosmology}\label{EBDE in BD}
The action of BD theory, in the canonical form, can be written
\cite{Arik}
\begin{equation}
 S=\int{
d^{4}x\sqrt{g}\left(-\frac{1}{8\omega}\phi ^2
{R}+\frac{1}{2}g^{\mu \nu}\partial_{\mu}\phi \partial_{\nu}\phi
+L_M \right)},\label{act1}
\end{equation}
where ${R}$ is the scalar curvature and $\phi$ is the BD scalar
field. Varying the above action with respect to FRW metric,
$g_{\mu\nu}$ and the BD scalar field $\phi$, we obtain the
following field equations
\begin{eqnarray}
 &&\frac{3}{4\omega}\phi^2\left(H^2+\frac{k}{a^2}\right)-\frac{1}{2}\dot{\phi} ^2+\frac{3}{2\omega}H
 \dot{\phi}\phi=\rho_m+\rho_D,\label{FE1}\\
 &&\frac{-1}{4\omega}\phi^2\left(2\frac{{\ddot{a}}}{a}+H^2+\frac{k}{a^2}\right)-\frac{1}{\omega}H \dot{\phi}\phi -\frac{1}{2\omega}
 \ddot{\phi}\phi-\frac{1}{2}\left(1+\frac{1}{\omega}\right)\dot{\phi}^2=p_D,\label{FE2}\\
 &&\ddot{\phi}+3H
 \dot{\phi}-\frac{3}{2\omega}\left(\frac{{\ddot{a}}}{a}+H^2+\frac{k}{a^2}\right)\phi=0,
 \label{FE3}
\end{eqnarray}
Hereafter, we consider the flat FRW universe, thus we set $k=0$.
At this point the system of our equations is not closed and we
still have a freedom to choice the scalar field. In the framework
of BD cosmology the BD scalar field $\phi$ is usually assumed to
has a power law relation in terms of scale factor, namely
\cite{riazinasr,riazi}
\begin{equation}\label{phipl}
\phi=\phi_0a(t)^{\varepsilon}.
\end{equation}
A case of particular interest is that when $\varepsilon$ is small
whereas $\omega$ is high so that the product $\varepsilon\omega$
results of order unity \cite{banpavon}. This is interesting
because local astronomical experiments set a very high lower bound
on $\omega$ \cite{Will}; in particular, the Cassini experiment
implies that $\omega
> 10^4$ \cite{bertotti,acqui}. Taking the time derivative of
relation (\ref{phipl}), we obtain
\begin{equation}\label{phidot}
 \frac{\dot{\phi}}{\phi}=\varepsilon \frac{\dot{a}}{a}=\varepsilon H.
\end{equation}
Using Eqs. (\ref{phipl}) and (\ref{phidot}), the first Friedmann
equation (\ref{FE1}) can be rewritten ($k=0$)
\begin{equation}\label{frid1}
  H^2(1-\frac{2\omega}{3}\varepsilon^2+2\varepsilon)=\frac{4\omega}{3\phi^2}(\rho_D+\rho_m).
\end{equation}
We introduce the fractional energy densities corresponding to each
energy component as
\begin{eqnarray}
\Omega_m&=&\frac{\rho_m}{\rho_{\mathrm{cr}}}=\frac{4\omega\rho_m}{3\phi^2
H^2}, \label{Omegam} \\
\Omega_D&=&\frac{\rho_D}{\rho_{\mathrm{cr}}}=\frac{4\omega\rho_D}{3\phi^2
H^2}, \label{OmegaD1}
\end{eqnarray}
where we have defined the critical energy density as
\begin{eqnarray}\label{rhocr}
\rho_{\mathrm{cr}}=\frac{3\phi^2 H^2}{4\omega}.
\end{eqnarray}
The reason for this definition comes from the fact that in BD
theory, the non-minimal coupling term $\phi^2 R$ replaces with the
Einstein-Hilbert term ${R}/{G}$ in such a way that
$G^{-1}_{\mathrm{eff}}={2\pi \phi^2}/{\omega}$, where
$G_{\mathrm{eff}}$ is the effective gravitational constant as long
as the dynamical scalar field $\phi$ varies slowly.

Using generalized ghost energy density (\ref{rhoD}) we can rewrite
Eq. (\ref{OmegaD1}) as
 \begin{eqnarray}\label{OmegaD2}
\Omega_{D} = \frac{4\omega}{3\phi^{2}H}\left( \alpha + \beta
H\right), \label{1}
\end{eqnarray}
Using definitions (\ref{Omegam}) and (\ref{OmegaD1}), Eq.
(\ref{frid1}) can be expressed as
\begin{eqnarray}
\Omega_{D}+\Omega_{m}= \gamma,\label{frid2}
\end{eqnarray}
where we have defined
\begin{eqnarray}
\gamma=1-\frac{2\omega}{3}\varepsilon^2+2\varepsilon.
\end{eqnarray}
Clearly for $\varepsilon=0$ ($\omega \rightarrow \infty$) we have
$\gamma=1$. In this case the BD scalar field becomes constant and
Einstein gravity is restored. Taking the time derivative of
Friedmann equation (\ref{frid1}) we find
\begin{equation}
2H \dot{H} \gamma
=\frac{4\omega}{3\phi^2}(\dot{\rho}_D+\dot{\rho}_m)-\frac{8\omega
\dot{\phi}}{3 \phi^3}(\rho_D+\rho_m).
\end{equation}
Using the continuity equations, as well as Eq. (\ref{phidot}) we
get
\begin{eqnarray}
2\dot{H}\gamma=-\frac{4\omega\rho_{D}}{\phi^{2}}\left(
1+u+\omega_{D}\right) -\frac{8\omega
\varepsilon}{3\phi^{2}}\rho_{D} \left( 1+u\right) .
\end{eqnarray}
Dividing by $H^2$, we have
\begin{eqnarray}
\frac{2\dot{H}\gamma}{H^{2}}=-3\Omega_{D}\left(
1+u+\omega_{D}\right) -2\varepsilon\Omega_{D}\left( 1+u\right).
\end{eqnarray}
Using the fact that $(1+u)\Omega_{D}={\gamma}$, we obtain
\begin{eqnarray}\label{dotH2}
\frac{\dot{H}}{H}=-H\left(
\varepsilon+\frac{3}{2}+\frac{3}{2}\frac{\Omega_{D}\omega_{D}}{\gamma}\right).
\end{eqnarray}
Finally for the time derivative of ghost energy density, we obtain
\begin{eqnarray}
\dot{\rho}_{D}=\left( \alpha+2\beta H\right) \dot{H}=-H^{2}\left(
\varepsilon+\frac{3}{2}+\frac{3}{2}\frac{\Omega_{D}\omega_{D}}{\gamma}\right)
\left( \alpha+2\beta H\right).
\end{eqnarray}
Inserting this relation in Eq. (\ref{consd}), it is a matter of
calculation to show that
\begin{eqnarray}
w_{D}= \frac{\gamma\left(  \frac{2}{3}\varepsilon-1\right)+
\frac{4\varepsilon\xi H\gamma}{3}
}{\left(2\gamma-\Omega_{D}\right) +{2\xi H}\left(
\gamma-\Omega_{D}\right) }.\label{wD2}
\end{eqnarray}
When $ \beta=0 $ one recovers
\begin{eqnarray}
w_{D} =\frac{\gamma}{\left(2\gamma-\Omega_{D}\right)  }\left(
\frac{2}{3}\varepsilon-1\right).
\end{eqnarray}
which is the EoS parameter of the original GDE in BD theory
presented in \cite{Ebr}. On the other hand, in the absence of BD
scalar field $\varepsilon=0 $ $(\gamma=1)$ we obtain the result of
the previous section, namely
\begin{eqnarray}
w_{D}=\frac{-1  }{\left(2-\Omega_{D}\right) +2 \xi H\left(
1-\Omega_{D}\right) }.
\end{eqnarray}
The solar-system experiments give the lower bound for the value of
$\omega$ to be $\omega > 40000$ \cite{bertotti}. However, when
probing the larger scales, the limit obtained will be weaker than
this result. It was shown \cite{acqui} that $\omega$ can be
smaller than 40000 on the cosmological scales. Also, Wu and Chen
\cite{wu} obtained the observational constraints on BD model in a
flat universe with cosmological constant and cold dark matter
using the latest WMAP and SDSS data. They found that within
$2\sigma$ range, the value of $\omega$ satisfies $\omega<-120.0$
or $\omega>97.8$ \cite{wu}. They also obtained the constraint on
the rate of change of $G$ at present
\begin{equation}\label{exp1}
    -1.75 \times 10^{-12} yr^{-1} <
\frac{\dot{G}}{G} < 1.05 \times 10^{-12} yr^{-1},
\end{equation}
at $2\sigma$ confidence level. As a result in our case with
assumption (\ref{phipl}) we get
\begin{equation}\label{exp2}
\frac{\dot{G}}{G}=\frac{\dot{\phi}}{\phi}=\varepsilon H< 1.05
\times 10^{-12} yr^{-1}.
\end{equation}
 This relation can be used to put an upper bound on
 $\varepsilon$. Assuming the present value of the Hubble
 parameter to be $H_{0}\simeq 0.7$, we obtain
 \begin{equation}\label{exp3}
 \varepsilon<0.01.
\end{equation}
The GGDE model in BD framework has an interesting feature compared
to the GDE model in Einstein's gravity. It was shown that in
standard cosmology based on Einstein's theory, the EoS parameter
of the noninteracting GGDE cannot cross the phantom line $w_D=-1$
and at the late time where $\Omega_D\rightarrow1$ approaches
$-1$\cite{shemov}. Choosing $\Omega_{D0}=0.72$, $H_{0}=0.7$ for
the present time and a suitable choice of $\xi$, this inequality
valid provided we take $\varepsilon<0.01$ in a narrow range which
is consistent with observations. This indicates that one can
generate a phantom-like EoS for the  GGDE in the BD framework. One
should note that increasing $\xi$ can exclude crossing the phantom
line which indicates a negative contribution of the subleading
term $H^2$ with respect to the leading term $H $ in the energy
density.

Since the dynamic of the universe should be discussed in term of
the effective EoS parameter thus in addition to the EoS parameter
of the GGDE we also study the effective EoS parameter, $w_{\rm
eff}$, which is defined as
\begin{equation}\label{wef}
w_{\rm eff}=\frac{P_t}{\rho_t}=\frac{P_D}{\rho_D+\rho_M},
\end{equation}
where $\rho_t$ and $P_t$ are, respectively, the total energy
density and total pressure of the universe. As usual we assumed
the dark matter is in the form of pressureless fluid ($P_M$=0).
Using relation (\ref{wD2}) for the flat case one can find
\begin{equation}\label{wef2}
w_{\rm eff}=\frac{\Omega_D}{\gamma}w_D=\frac{\Omega_{D}\left(
\frac{2}{3}\varepsilon-1\right)+ \frac{4\varepsilon\xi
H\Omega_{D}}{3}  }{\left(2\gamma-\Omega_{D}\right) +2 \xi H\left(
\gamma-\Omega_{D}\right) }.
\end{equation}

\begin{figure}\epsfysize=5cm
{ \epsfbox{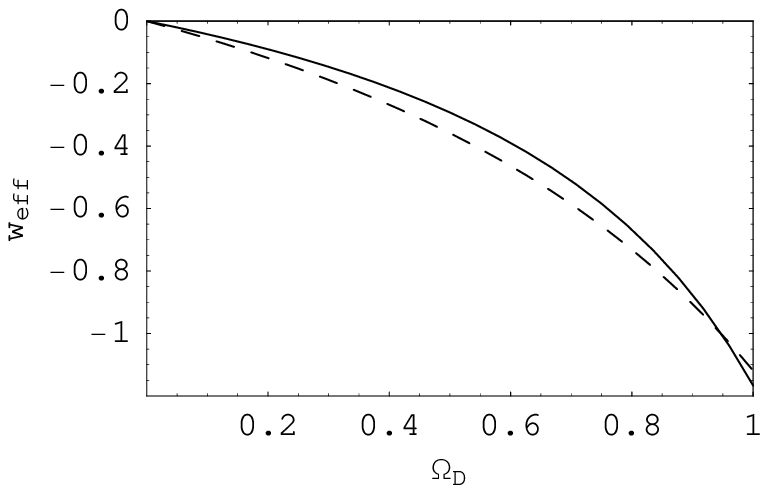}}\epsfysize=5cm { \epsfbox{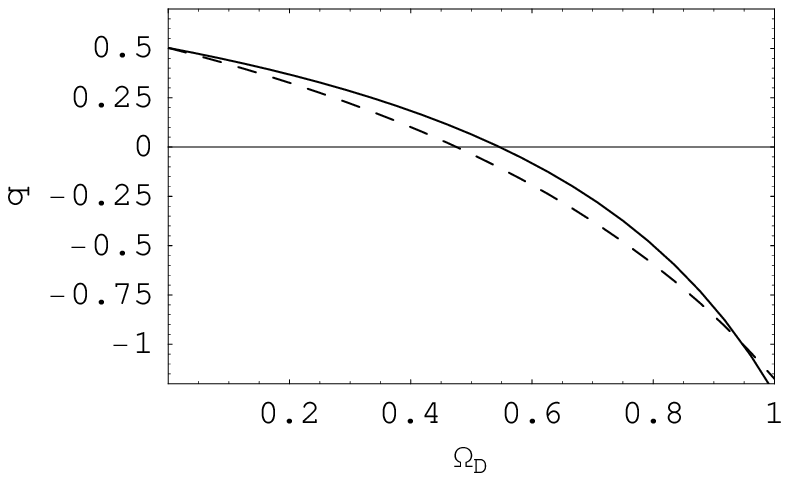}}\caption{In
these figures $w_{\rm eff}$ and $q$ for GGDE and GDE are plotted
against $\Omega_D$. Solid curve corresponds to GGDE and dashed
ones belong to GDE. In both of these figures we take
$\omega=10000$,$\varepsilon=0.003$, $H_0=0.7$, $\xi=0.5$.}
\label{i1}
\end{figure}
It is also interesting to study the behavior of the deceleration
parameter defined as
\begin{eqnarray}\label{qb}
q=-\frac{\ddot{a}}{aH^2}=-1-\frac{\dot{H}}{H^2}.
\end{eqnarray}
Substituting Eq. (\ref{dotH}) in the above relation one can easily
reach
\begin{equation}\label{qb2}
q=\frac{1}{2}+\varepsilon+\frac{3}{2}\frac{\Omega_D w_D}{\gamma}.
\end{equation}
Inserting Eq. (\ref{wD2}) into  (\ref{qb2}) yields
\begin{equation}\label{qb3}
q=\frac{1}{2}+\varepsilon+ \frac{3}{2}
\frac{\Omega_D(\frac{2}{3}\varepsilon-1)+\frac{4\varepsilon\xi H
\Omega_D}{3}}{(2\gamma-\Omega_D)+ 2 \xi H(\gamma-\Omega_D)}.
\end{equation}
When $ \beta=0 $ we obtain \cite{Ebr}
\begin{eqnarray}\label{qb4}
q=\frac{1}{2}+\varepsilon +\frac{3}{2}\frac{\Omega_{D}\left(
\frac{2}{3}\varepsilon-1\right)
}{\left(2\gamma-\Omega_{D}\right). }\label{qQ}
\end{eqnarray}
In the limiting case $\varepsilon=0$ ($\omega\rightarrow\infty$)
we have $\gamma=1$ and hence the BD scalar field becomes trivial;
as a result Eq. (\ref{qb3}) reduces to its respective expression
in flat standard cosmology obtained in the previous section
\begin{eqnarray}
q=\frac{1}{2}- \frac{3}{2} \Omega_D\left[2-\Omega_D+ 2\xi
H(1-\Omega_D)\right]^{-1}.
\end{eqnarray}
Let us study some special cases of interest for the deceleration
parameter $q$. If we take $\Omega_{D0}=0.72$, $H_{0}\simeq 0.7$
and  for the present time and choosing $\varepsilon=0.002,
\xi=0.1$ and $\omega=10000$ we obtain $q_0=-0.35$, which is
consistent with the present value of the deceleration parameter
obtained in \cite{daly}. Transition from deceleration to
acceleration occurs at $\Omega_D=0.46$. It is worth noting that
GGDE results in a smaller rate of acceleration in comparison with
the GDE then this models leads a delay in different epoches of the
cosmic evolution with respect to the original GDE (for instance
the GDE with choice of a same set of parameters lead $q_0=-0.37$).
For a better insight about these two models we plotted $w_{eff}$
and $q$ for both of models in Fig\ref{i1}.

Finally, we obtain a differential equation governing the evolution
of GGDE from early deceleration to the late time acceleration.
Following the method of the previous section we find
\begin{eqnarray}
\dot{\Omega}_{D}&&=\frac{\alpha\Omega_{D}H }{\alpha+\beta H}\left(
q+1\right)  -2\Omega_{D}\varepsilon H\\&&=\Omega_{D}H\left(
\frac{1+q}{1+\xi H} -2\varepsilon\right).
\end{eqnarray}
Inserting $q$ from (\ref{qb3}) and using relation
$\dot{\Omega}_D=H\Omega^{\prime}_D$, we obtain
\begin{eqnarray}\label{OmegaDev}
\Omega^{\prime}_D=\Omega_D \left(\frac{1+q}{1+\xi
H}-2\varepsilon\right),\label{omega prime}
\end{eqnarray}
where the prime denotes derivative with respect to $x=\ln a$ and
$q$ is given by Eq. (\ref{qb3}). In the limiting case $\beta=0$
one recovers the result obtained in \cite{Ebr}.

\section{Closing remarks}\label{sum}
A phenomenological GGDE model whose energy density is of the form
$\rho=\alpha H+\beta H^{2}$ was recently proposed to explain the
observed acceleration of the universe expansion. This model
originates from the fact that the vacuum energy of the Veneziano
ghost field in $QCD$ is of the form, $H +O(H^2)$. It was shown
that the difference between the vacuum energy of quantum fields in
Minkowski space and in FRW universe can play the role of observed
dark energy.

In this paper we first studied the cosmological implications of
the GGDE model in standard cosmology. We found that in this model,
the universe approaches to a de Sitter phase at late times. Then,
we extended our study to the BD cosmology. We can classify our
achievements in two categories. The first is that in the BD
framework the GGDE can cross the phantom line with suitable choice
of the free parameters while in the standard cosmology we found a
de Sitter phase as the fate of the universe. The second result is
that the subleading term $H^2$ can lead a delay in different
epoches of the cosmic evolution. We also discussed this result
explicitly by a numerical evaluation and showed that taking a same
set of parameters result $q_0=-0.35$ for the GGDE while for GDE
gives $q_0=-0.37$. It is easily seen from Fig.\ref{i1} that GDE
(dashed curve) enters the acceleration phase sooner than the GGDE
(solid curve). This point was also addressed in standard cosmology
\cite{CaiGhost2} as the negative contribution of the subleading
term $H^2$ in the energy density.

Finally, we would like to mention that in the present work we only
considered the mathematical presentation of the GGDE model in the
framework of Brans-Dicke cosmology. Other aspects of this model
will be addressed in the separate works. For example, recently, we
have studied the stability of the GGDE model against perturbations
in the FRW background. Based on the square sound speed analysis,
due to the existence of a free parameter in this model, the GGDE
model is theoretically capable to lead a dark energy dominated
stable universe \cite{Stab}. The extension of this study to the
Brans-Dicke cosmology is under investigation and will be addressed
elsewhere.

\acknowledgments{We are grateful to the referee for constructive
comments which helped us to improve the paper significantly. A.
Sheykhi thanks Shiraz University Research Council. This work has
been supported financially by Research Institute for Astronomy and
Astrophysics of Maragha (RIAAM), Iran.}

\end{document}